Quantum-like perception entanglement leads to advantageous collective decisions

David Lusseau[1]

[1]Institute of Biological and Environmental Sciences, University of Aberdeen, Aberdeen, AB24 2TZ, UK. E-mail: d.lusseau@abdn.ac.uk

**ABSTRACT:** Social animals have to make collective decisions on a daily basis. In most instances, these decisions are taken by consensus, when the group does what the majority of individuals want. Individuals have to base these decisions on the information they perceive from their socioecological landscape. The perception mechanisms they use can influence the cost of collective decisions. Here I show that when group-living individuals perceive their environment concurrently for the same decisions, a quantum-like perception entanglement process can confer less costly collective decisions than when individuals collect their information independently. This highlights a mechanism that can help explain what may seem to be irrational group-living behavior and opens avenues to develop empirical tests for quantum decision theory.

Social animals, including humans, make collective decisions for many facets of their daily lives *(1, 2)*. Collective decisions have evolved in group-living species because they appear to be less costly than personal decisions when individuals try to maintain group membership. However, some observed features of collective decisions can seem irrational. For example, group decisions appear to be more accurate as group size increases, even when individuals are uninformed *(2)*. A number of models have been developed to try and explain emergent collective behavior as well as infer the mechanisms through which collective decisions occur *(2-4)*. In all cases, we assume that individuals collect information, in some form or other, to make a decision *(1)*. However, the influence of perception systems on this process has received little attention to date.

Before making a decision, individuals have to perceive the landscape in which they find themselves *(5)*. In short, individuals have to generate knowledge from their socioecological landscape by first acquiring information from data (the landscape, Figure 1a) *(6, 7)*. This mechanism involves two error processes: when the individual generates information from data and when the individual generates knowledge from the information it receives. Perception mechanisms are involved in generating information and therefore play a critical, yet unassessed, role in collective decisions. Importantly, one advantage of group living is shared perception. That is, the reliance of individuals on others to be vigilant *(8)* or find resources such as food *(9)*. When individuals rely on others for perception, the knowledge generation process described above will co-vary between individuals within a group (Figure 1a). Eavesdropping, when one individual acquires information from the knowledge generation process of others, is a process that can drive this covariance *(5, 9)*.

Eavesdropping is thought to be prevalent in echolocating mammals *(10-13)*. In such instances, silent individuals can receive the same information as echolocating conspecifics by listening to the echoes from the calls those produced. They can then use this information for a range of behavioral decisions *(11, 12)*. However, this process is not exactly eavesdropping (Figure 1b). What we have instead is an entanglement in the information generation process between individuals present together in a group. This perception entanglement is an extreme case of perception covariance between group members which offers an interesting avenue to assess how perception mechanisms might affect collective decisions.

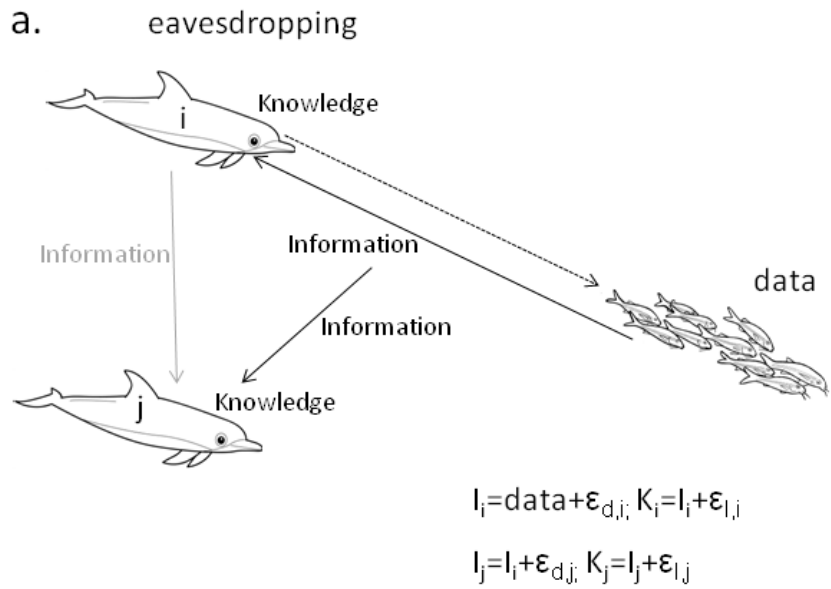

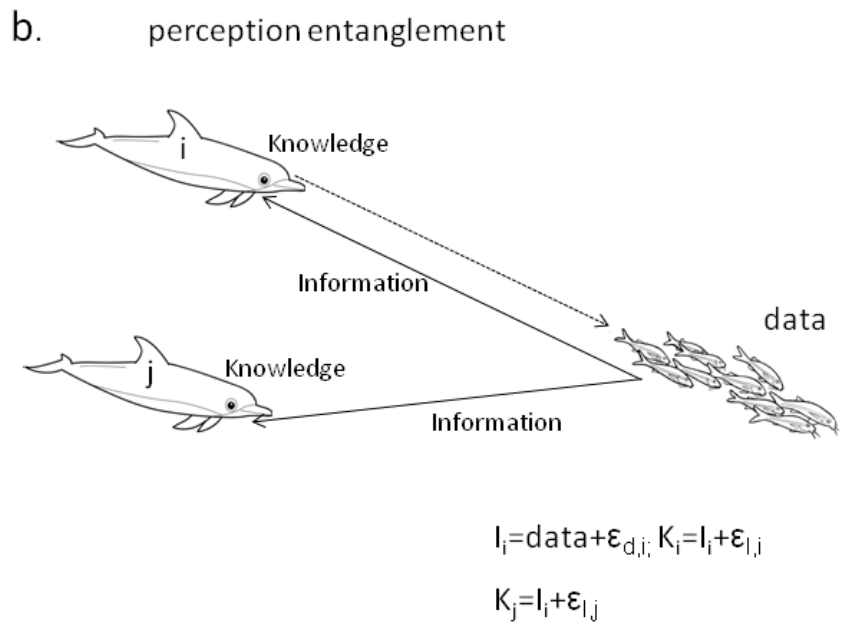

**Figure 1.** Two ways in which individuals can derive knowledge from their environment: in all cases one individual, i, generate information, I, with some error, $\varepsilon_d$, from data (its environment) and then generate knowledge, K, from I with some error, $\varepsilon_I$. Another individual can derive knowledge from the sampling of individual i. **(a)** It, j, can eavesdrop on i and generate information, $I_j$, either from $I_i$, or from observing individual i (gray arrow, for example, seeing i flee from an area). **(b)** When i uses an active perception system, such as echolocation (dotted line), to generate $I_i$, then j can also receive $I_i$ directly but interpret it to generate $K_j$, with its own error process, $\varepsilon_{d,j}$. the perception of these two individuals is then entangled with consequences for the knowledge they generate.

We know that, for echolocating species, individuals in groups regularly listen to the echoes of others *(14)*. Echolocation seems then irrational because echolocating individuals confer advantages to their closest competitors. They provide information about their environment to others who can then use it to their advantage and to the detriment of the echolocation producer. Some inclusive fitness processes *(15)* or simple cooperative mechanisms *(16)* could explain why this perception mechanism evolved. However, echolocation, and eavesdropping on echoes, is present in populations where grouping is not influenced by genetic relationships *(10)*, interaction rates are too low for cooperation to be advantageous *(13)*, or even in mix-species groups *(10)*.

Here I develop three contrasting mechanistic models of perception influences on collective decisions. I aim to resolve the "group-echolocation" conundrum, as an extreme case of perception co-variance, by testing whether mechanisms leading to such perception entanglement could yield beneficial collective decisions. One type of decisions groups have to regularly take collectively is the timing of their activities. Conradt and Roper *(3)* presented a simple and elegant model of timing decisions to show that consensus-building should be, in most cases, the most beneficial way to make such collective decisions. Here I extend their model to account for the effect of perception during individual's knowledge acquisition to make decisions about activity timing. First, we assess whether benefits could emerge from perception interference; that is interference in knowledge generation between individuals that have many common experiences *(17)*. Secondly, we assessed the costs of collective decisions when individuals make a rational, probabilistic, choice between generating knowledge from their own information or the information generated by others in their group *(17)*. Lastly, we test whether a form of entanglement could result in beneficial collective decisions.

Entanglement is a process that has been formally developed and explored in quantum mechanics to explain observed departures from expectations in classical mechanics *(18, 19)*. Recently, formal and informal applications of quantum processes have been used to try and explain departure from rationality in game and decision theories. Informal applications simply borrow the mathematics and concepts behind known quantum effects *(20, 21)*, as opposed to look for quantum effects in cognition *(17)*, and that avenue is proving fruitful. I apply this approach here. I assume that individuals can be in two 'perception' states that are superposed: they can either use personal information (PI) or information generated by others (OI). Individuals do not choose a state in a rational manner (which reduces to a probabilistic approach, Figure 2a) but only select a quantum state in this Hilbert state space when they sample it (when they need to generate knowledge, Figure 2a). As I described above, all individuals in the group have to go through this process based on entangled perception. This means that the state selection process is entangled between individuals.

I simulated group decisions using these three perception mechanisms and contrasted them to the original model in which individuals always use PI (Figure 2a). As error and group size can influence the benefits of collective decisions *(3)*, I simulated decisions over a range of error size ($\sigma^2$, Eq.2) and group size (*n*, Eq. 1). Social structure can influence the evolution of cooperation and the genetic structure of populations. I therefore also assessed whether structuring in interactions could influence these benefits under the three perception models. I sampled groups at random from social networks composed of units that interacted at varying rate. To do so, I simply derived random social networks with modularity ranging from 0.02, in which case these units are spurious, to 0.7 (strong structuring) using the method described in *(16)*. I present here results for a population of 100 individuals with three social units. I run simulations for a range of population size and number of social units but this did not affect the results presented below. I simulated 100 decisions for each combination of error size, group size and network modularity. Groups were selected from the social network by first selecting at random a seed individual and then selecting the remaining group members using a random draw weighted by the association rate indices of the seed individual. The costs of collective decisions were derived from extension from the Conradt-Roper model:

$$cost_{CR} = \sum_i \left| t_{real,i} - t_{perceived,\frac{n+1}{2}} \right| \qquad [1]$$

Which is the sum over all individuals, *i*, of the absolute difference in the real optimal timing of the activity, $t_{real}$, for that individual and the perceived optimal timing of the activity, $t_{perceived}$, for the median individual ($\frac{n+1}{2}$) where

$t_{perceived,i} = t_{real,i} + \varepsilon$ and $\varepsilon \sim N(0,\sigma^2)$  [2]

I focus on this perception error. The cost computation does not change, only the way $t_{perceived}$ is inferred. For simplicity I randomly drew $t_{real}$ for each individual from a log-normal distribution with mean log(55min) and standard deviation 0.3.

We first assume that perception interference from common experiences influence the actual error process. In that case, the error for an individual will be correlated with the error of others depending on its association rate indices with group members:

$t_{perceivedcor,i} = t_{real,i} + \varepsilon * chol[\mathbf{AI}_{n,n}]$  [3]

where $chol[\mathbf{AI}_{n,n}]$ is the Cholesky decomposition of the subset of **AI** for the n individuals in the group.

In the case of a classical interaction of information, where individuals make a rational choice between OI and PI, the perceived timing is itself correlated between individuals:

$t_{perceivedinter,i} = t_{perceived,i} * chol[\mathbf{AI}_{n,n}]$  [4]

In both those models association rate indices are used as proxy of interference and interaction because close associates are also more likely to be physically closer to each other within a group *(22)* and individuals can be discriminated from their echoes in, at least, some species *(23)*.

The perceived timing of activity under entanglement can be inferred using an iterative-game like approach. We defined the entanglement as an iterative 2-player game-like interaction (Figure 2b) played over all possible pairs in the group *(18)*. The perception mode on which both individuals (*i* and *j*) decide after an interaction, $|d_{i,f}\rangle$ and $|d_{j,f}\rangle$, are defined as:

$$|d_{i,f}\rangle = \mathbf{J}^{-1^{\otimes 2}} \mathbf{U}_i \otimes \mathbf{U}_j \mathbf{J}^{\otimes 2} |d_{i,0}\rangle$$  [5]

$$|d_{j,f}\rangle = \mathbf{J}^{-1^{\otimes 2}} \mathbf{U}_j \otimes \mathbf{U}_i \mathbf{J}^{\otimes 2} |d_{j,0}\rangle$$  [6]

Each player received a qubit at the beginning of the game and start with a propensity for PI. **J** is the entanglement operator defined by the entanglement assumed between the two players based on their association rate index, $\mathbf{AI}_{ij}$:

$$\mathbf{J}_{ij} = \mathbf{I}\cos\frac{\frac{\pi}{2}\mathbf{AI}_{ij}}{2} + i\hat{\sigma}_x(\sin\frac{\frac{\pi}{2}\mathbf{AI}_{ij}}{2})$$  [7]

So that the entanglement rate is maximised when $\mathbf{AI}_{ij} = 1$. $\hat{\sigma}_x$ is the bit-flip operator $\begin{bmatrix} 0 & 1 \\ 1 & 0 \end{bmatrix}$ *(21)*.
and **U** is the unitary matrix defined by the Hilbert space of quantum states *(18)*. At this stage, the two qubits, one for each player, can be assigned to four possible combinations $|PI,PI\rangle, |PI,OI\rangle, |OI,PI\rangle, |OI,OI\rangle$. Here we simplified the unitary operator so that only the phase, θ, can vary between the two players (Figure 2a) so that for individual *j*:

$$\mathbf{U}_j = \begin{bmatrix} e^{-i\pi} \cos\left(\frac{\theta_j}{2}\right) & e^{i\pi} \sin\left(\frac{\theta_j}{2}\right) \\ -e^{-i\pi} \sin\left(\frac{\theta_j}{2}\right) & e^{i\pi} \cos\left(\frac{\theta_j}{2}\right) \end{bmatrix} \quad [8]$$

and $\theta_j$ is drawn randomly and uniformly from [0;π] at every decision *(18)*. The final timing decision of individual *i* is simply:

$$t_{perceivedquant,i} = \Sigma \begin{bmatrix} t_{perceived,i} \\ t_{perceived,i} \\ t_{perceived,j} \\ t_{perceived,j} \end{bmatrix} \times \left|\left| d_{i,f} \right\rangle\right|^2 \quad [9]$$

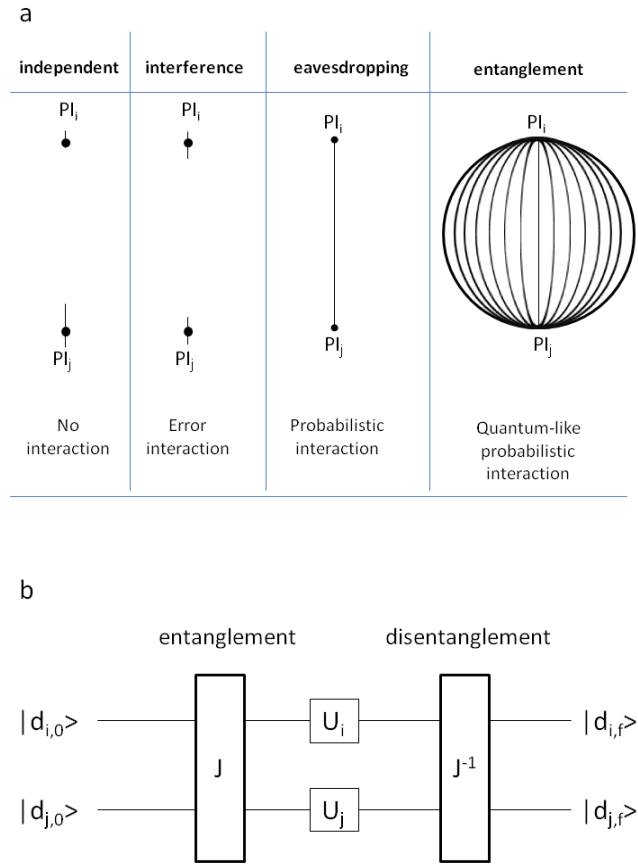

**Figure 2**. **(a)** A graphical analogy of the three perception models tested in these simulations for two individuals, *i* and *j*, compared to the current collective decision model where individuals always use PI ('independent'). Here $PI_j=OI_i$ and $PI_i=OI_j$. Perception interference means that the error (error bars) around PI are correlated between *i* and *j*. When *j* eavesdrops on *i*, it integrates $PI_i$ and $PI_j$ to make a decision and chooses a state somewhere in between. When the perception of *i* and *j* are entangled, the choice is similar but can be approached from a number of angles depending on a phase, θ, which is a feature of the decision's wave function *(20)*. The choice is only known when it is sampled by the individual (when the individual needs to generate knowledge). **(b)** A circuit diagram representing this process. The entangled initial state results from entangling *i* and *j* with **J** while both had an initial state, $\left|d_{j,0}\right\rangle$ for *j*, of using 'PI' (and its associated knowledge of $t_{perceived}$). Entangled *i* and *j* then choose their state ('PI' or 'OI' using $U_i$ and $U_j$ respectively) and knowledge, $t_{perceivedquant}$, is estimated from the decision that was taken after individuals become disentangled $\left|d_{j,f}\right\rangle$.

In all cases we estimated the costs to all individuals of classical consensus collective decisions (Eq. 1). I applied the three perception models to the same set of simulations so that I could compare their cost by simply taking the difference in costs resulting from each perception model (Eqs. 3,4,9) and the classical one (Eq. 2) for all iterations. I found that both perception interference (Figure 3a) and rational use of PI and OI (Figure 3b) do not lead to collective decisions that are less costly than decisions taken using PI only. However, we find that for a wide range of error size and group size conditions, decisions taken with perception entanglement are more beneficial than those using PI only (Figure 3c). We are therefore in the counter-intuitive situation that sharing the information acquisition process is less costly to individuals than acquiring their own information. Social network modularity did not influence the difference in costs (Figure 3d). Hence, the observed benefits of perception entanglement should exist across a wide range of social structure. This finding converges with the observations we previously described that echolocation occurs over a range of social landscapes and contexts.

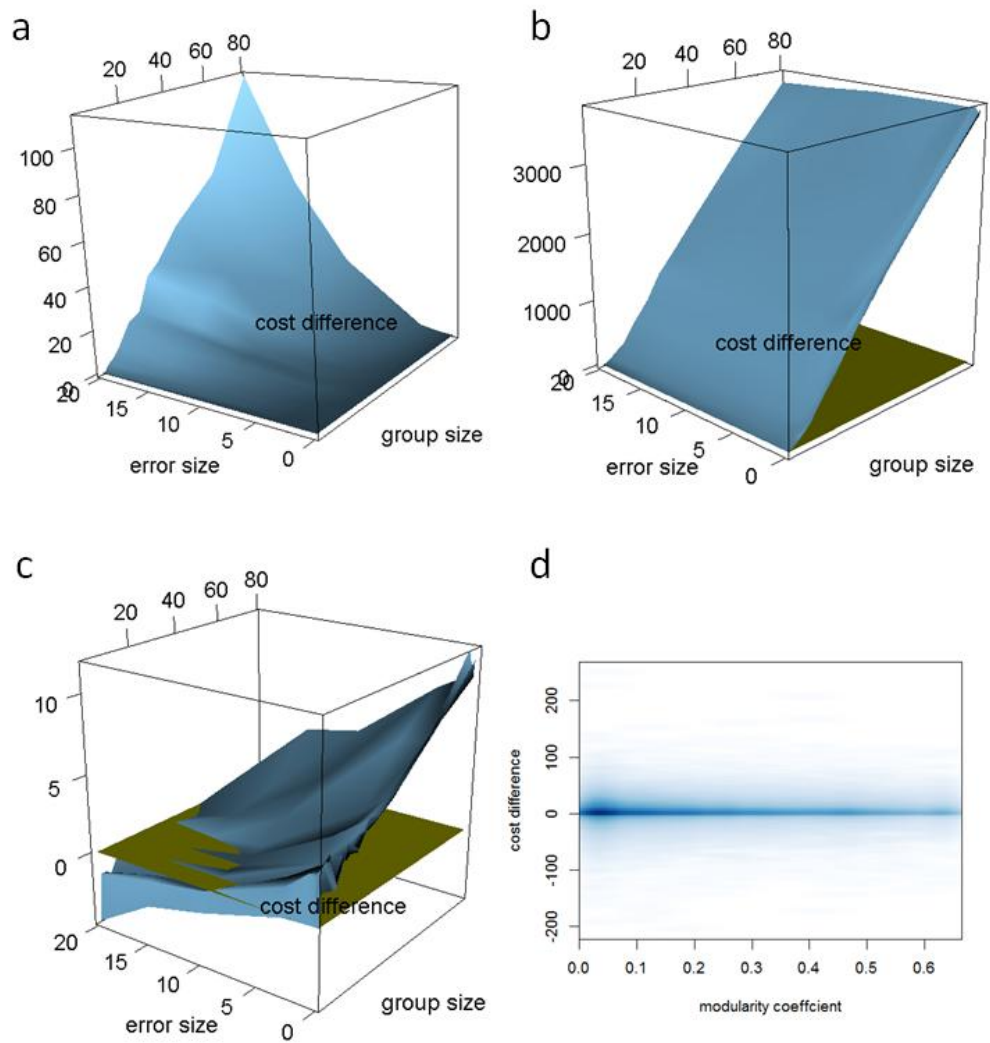

**Figure 3**. The estimated costs of simulated decisions show that **(a)** the cost of collective decisions under perception interference is always greater than if individuals perceive the environment independently (cost difference>0 for all simulated decisions). Outcomes are similar for perception interactions **(b)**. However, decisions taken under perception entanglement **(c)** can be less costly than independent sampling (cost difference <0) for medium to large errors, $\sigma^2$, depending on group size, *n*. Surfaces are the median difference in costs and the zero-cost plane is emphasized. The social network modularity did not influence the difference in costs when comparing perception entanglement and independent sampling (**d**, smoothed density of scatterplot using `smoothScatter` in R 2.14).

We now realize that rational choice models are limited in their abilities to predict decisions and actions *(24-26)*. Quantum game theory *(20)* and quantum decision theory (QDT) *(17)* are emerging as alternative approaches to at least imitate observed irrational behaviors and choices. These are not based on quantum effects as we understand them in mechanics and electromagnetism, but they borrow the concept of the Schrödinger equation to describe the 'fuzzy' mechanisms that takes place when individuals take decisions. Considering decision making as quantum measurements has been proposed for a long time *(24)*, however empirical tests have been lagging *(18)*. Perception plays a key role in all facets of behavioral decisions and here it seems that a quantum-like model of perception can explain the seemingly irrational emergence of echolocation in groups. This active form of perception can emerge as a beneficial component of a collective, entangled, decision-making process when individuals exist in superposed states of information acquisition. Here we used a simple assumption to define the entanglement rate between individuals. Further work is required to develop formal representations of perception entanglement. For this our model bridges with recent developments of quantum cognition models *(17)*. This concept can also be extended to account for other form of interactions during perception such as quantum interference and importantly points at a study system in which empirical tests of QDT can be developed. Interference and entanglement are emerging as important features of decision-making processes yielding what seem to be irrational behaviors under classical rational choice models *(24)*.

**References and notes**

1. L. Conradt, T. J. Roper, Consensus decision making in animals. *Trends in Ecology & Evolution.* **20**, 449-456 (2005).

2. I. D. Couzin *et al.*, Uninformed Individuals Promote Democratic Consensus in Animal Groups. *Science.* **334**, 1578-1580 (2011).

3. L. Conradt, T. J. Roper, Group decision-making in animals. *Nature.* **421**, 155-158 (2003).

4. I. D. Couzin, J. Krause, N. R. Franks, S. A. Levin, Effective leadership and decision-making in animal groups on the move. *Nature.* **433**, 513-516 (2005).

5. E. Danchin, L. A. Giraldeau, T. J. Valone, R. H. Wagner, Public information: from nosy neighbors to cultural evolution. *Science.* **305**, 487-491 (2004).

6. K. Körding, Decision Theory: What "Should" the Nervous System Do? *Science.* **318**, 606-610 (2007).

7. L. Barrett, S. P. Henzi, D. Lusseau, Taking sociality seriously: the structure of multi-dimensional social networks as a source of information for individuals. *Philosophical Transactions of the Royal Society B: Biological Sciences.* **367**, 2108-2118 (2012).

8. J. J. Templeton, L. A. Giraldeau, Vicarious sampling: The use of personal and public information by starlings foraging in a simple patchy environment. *Behav. Ecol. Sociobiol.* **38**, 105-114 (1996).

9. S. R. X. Dall, L. A. Giraldeau, O. Olsson, J. M. McNamara, D. W. Stephens, Information and its use by animals in evolutionary ecology. *Trends in Ecology & Evolution.* **20**, 187-193 (2005).

10. M. B. Fenton, Eavesdropping on the echolocation and social calls of bats. *Mamm. Rev.* **33**, 193-204 (2003).

11. T. Götz, U. K. Verfuß, H. Schnitzler, 'Eavesdropping' in wild rough-toothed dolphins (Steno bredanensis)? *Biology Letters.* **2**, 5-7 (2006).

12. G. Jones, Sensory Ecology: Echolocation Calls Are Used for Communication. *Current Biology.* **18**, R34-R35 (2008).

13. S. M. Dawson, Clicks and Communication: The Behavioural and Social Contexts of Hector's Dolphin Vocalizations. *Ethology.* **88**, 265-276 (1991).

14. L. G. Barrett-Lennard, J. K. B. Ford, K. A. Heise, The mixed blessing of echolocation: Differences in sonar use by fish-eating and mammal-eating killer whales. *Anim. Behav.* **51**, 553-565 (1996).

15. P. D. Taylor, G. Wild, A. Gardner, Direct fitness or inclusive fitness: how shall we model kin selection? *J. Evol. Biol.* **20**, 301-309 (2007).

16. M. Marcoux, D. Lusseau, Network modularity promotes cooperation. *J. Theor. Biol.* **324**, 103-108 (2013).

17. E. M. Pothos, J. R. Busemeyer, Can quantum probability provide a new direction for cognitive modeling? *Behav. Brain Sci.* **36**, 255 (2013).

18. K. Chen, T. Hogg, How Well Do People Play a Quantum Prisoner's Dilemma? *Quantum Information Processing.* **5**, 43-67 (2006).

19. J. Du, H. Li, X. Xu, X. Zhou, R. Han, Entanglement enhanced multiplayer quantum games. *Physics Letters A.* **302**, 229-233 (2002).

20. J. - Eisert, Quantum games and quantum strategies. - *Physical Review Letters.* **83**, 3077-3080 (1999).

21. A. P. Flitney, D. Abbott, An introduction to quantum game theory. *Fluct. Noise Lett.* **2**, R175-R187 (2002).

22. E. C. G. Owen, R. S. Wells, S. Hofmann, Ranging and association patterns of paired and unpaired adult male Atlantic bottlenose dolphins, Tursiops truncatus, in Sarasota, Florida, provide no evidence for alternative male strategies. *Can. J. Zool. -Rev. can. Zool.* **80**, 2072-2089 (2002).

23. M. Knörnschild, K. Jung, M. Nagy, M. Metz, E. Kalko, Bat echolocation calls facilitate social communication. *Proceedings of the Royal Society B: Biological Sciences.* **279**, 4827-4835 (2012).

24. V. I. Yukalov, D. Sornette, Decision theory with prospect interference and entanglement. *Theory Decis.* **70**, 283-328 (2011).

25. V. I. Yukalov, D. Sornette, Quantum decision theory as quantum theory of measurement. *Physics Letters A.* **372**, 6867-6871 (2008).

26. E. M. Pothos, J. R. Busemeyer, A quantum probability explanation for violations of 'rational' decision theory. *Proceedings of the Royal Society B: Biological Sciences.* **276**, 2171-2178 (2009).

**ACKNOWLEDGEMENTS**: This work received funding from the MASTS pooling initiative (the Marine Alliance for Science and Technology for Scotland) and their support is gratefully acknowledged. MASTS is funded by the Scottish Funding Council (grant reference HR09011) and contributing institutions.